# A Friend Recommendation System using Semantic Based KNN Algorithm

Srikantaiah K C*, Salony Mewara,  Sneha Goyal, Subhiksha S

*Department of Computer Science and Engineering,
SJB Institute of Technology, (Affiliated to Visvesvaraya Technological University),
Bengaluru, Karnataka, India.*
srikantaiahkc@gmail.com,
salonymewara@gmail.com, sneha.bwgs26@gmail.com, subhikshasarode@gmail.com

**Abstract:** *Social networking has become a major part of all our lives and we depend on it for day to day purposes. It is a medium that is used by people all around the world even in the smallest of towns. Its main purpose is to promote and aid communication between people. Social networks, such as Facebook, Twitter etc. were created for the sole purpose of helping individuals communicate about anything with each other. These networks are becoming an important and also contemporary method to make friends from any part of this world. These new friends can communicate through any form of social media. Recommendation systems exist in all the social networks which aid users to find new friends and unite to more people and form associations and alliances with people. With friends, there comes a robust friend recommendation system too. Graph based friend recommendation system are typically in use and are not the most accurate and holds the potential to drive users to wrong track or connect with people who could take advantage of the users such as fraud or scams. So we have tried to implement a recommendation system which is truer as compared to other models by adding some extra layers of character breakdown and user behavior. Since a vast amount of data exists which consist of user information, our main goal is to figure out each user's personality traits and conduct which will be used to help him/her finding out new users with same temperament. The experimental results show that our model gives 90% accuracy.*

**Keywords:** Friend Recommendation System, KNN Algorithm, Machine Learning, Recommendation System, Social Networks,

## 1. INTRODUCTION

People like making new friends and they might want to make new friends on various social media accounts so as to expand their social influences and talk to people in any part of the world by using online social networks, For example, business page proprietors on any website may want to influence as many people as possible for profit oriented advantages. The friend recommendations in the existing systems are based on the social interactions and personal information profiles of the user or the method by which a friend of a friend is referred to the user. A similar strategy is used on Facebook, Twitter and various other social networks where the recommended results are based on common factors such as mutual friends, age groups, terrestrial location, place of work and other such information, etc. However, there are some shortcomings to increasing the number of friends with this method. Friends recommended based solely on the number of friends they share may not necessarily have the same interests and habits as the target users. At the same time, this method can scrap a lot of useful and beneficial information on social networks, through which we can get more accurate and meaningful recommendation results from friends.

In this paper, we present recommendations for new friends based on the semantics in social network system. It recommends friends to users based on their lifestyle. By taking advantage of sensor-rich smartphones, it attempts to identify users' lifestyles based on user-centric sensor data, measures the similarity of lifestyles between users, and recommends friends to users based on the likelihood and similarities of lifestyles. We have made use of the KNN Algorithm, which is one of the easiest algorithm to recommend friends. It is termed as lazy learner as it does not immediately read the statistics and implement it. By using KNN algorithm we try to locate the users who have similar taste or interests as another user by choosing the books that could have been read by the users. Smartphones with sensors provide us with an edge and hence we can develop this model by suggesting friends based on the user similarity and features. We also put forward a Similarity metric to measure the similarity between different users and calculate the consequences of users in relation to lifestyle. In our recommendation model, we have tried to establish a friend recommendation mechanism for social networking websites by relating content attractions and finding out the users that can be recommended by using books that have been read. We picked books B1-B10 and got a good accuracy or edition to justify the process. Upon receipt of a request, our Friend Recommendation system returns a quote made up of people with the highest recommendation scores for that of the query user. Implementation of this Friend Recommendation system can be done on any smart phone preferably an android phone or any laptop and hence can be evaluated for its performance and accuracy measures.

The remaining paper is organized as follows: Section 2, provides a brief literature survey of friend recommendations. In Section 3, we describe in detail the problem and the components of system architecture In Section 4, we describe algorithms and methods for recommendation of friends based on the books read. The discussion of empirical findings and results is described in Section 5, Section 6 concludes and discusses the field of work in the future.

## 2. RELATED WORKS

Zhibo Wang, *et al.,* [1] propose a method called "Friendbook: A Semantically Based Friend Referral System for Social Networking", in which the recommendation of friends to users based on their lifestyle is done using the LDA algorithm. Social graphs are not weighted here. By utilizing devices that support sensors, user lifestyles can be extracted from user-centered sensor data from Friendbook, which recommends users to friends when their lifestyles have a high association between lifestyles between users and measure the similarity of lifestyles between users. It prototypes the everyday life of a user as life documents inspired by using text mining techniques. To further improve the recommendation accuracy feedback mechanism is assimilated by Friendbook, Shangrong Huang, *et al.,* [2] proposed Social Friend Recommendation Based on Multiple Network Correlation, which makes use of NC Based SFR algorithm to get good precision and highest accuracy. Two major components are characterized by NC-based SFR approach are: 1) by selecting important features from each system, related networks are associated and 2) before and after network alignment the network structure should be extremely preserved.

Jiang, *et al.,* [3] suggested a "social recommendation across multiple relational domains" that allows us to perceive a hybrid random walk method on a Star Structured Graph and also the RDR algorithm, which is organized in such a way that it helps to alleviate the main problems in a single domain by shifting knowledge from other domains and considering how social networks with multiple relational ones Provinces can be represented. Liu, et al., [4] implements many ways of optimization techniques in global and local structure preservation for the feature selection, which integrates both the global pairwise sample similarity and the preservation of the feature selection structure. Bian *et al.,* suggested a recommendation for online friends through collaborative filtering and personality matching based on personality matches. The aim of Matchmaker is to create recommendations from friends based on extensive contextual data from the interactions in the physical world of people and using the mutual understanding and social information between people in existing social network connections. In order to suggest friends to users who were chosen according to their personality, the best matchmaker uses relationships in the TV programs as a parallel comparison matrix. The ranking scheme of the system enables a gradual improvement of the consensus on the personality match. The social network can also lead to higher consumption of TV content by arousing the curiosity of users about the ranking process. Kwon and Kim [6] introduced a friend recommendation system that mainly uses the physical and social context and summarizes similar functions of multiple users and labels. The user with adaptive recommendations from massive information is provided by the context-aware systems. To recommend truly esteemed friends using context is the essential aspect of social computing technologies.

Naruchitparames,  *et al.,*[7] suggested recommendation systems to help improve the user experience using complex network theory and cognitive theory to provide post quality friend referrals while finding an individual's insight into friendship. The difficulty in developing recommendation mechanisms is mainly due to the diversity of social networks. Their research shows that combining genetic algorithms and network topology can produce superior recommendations compared to any single counterpart. This approach was tested on a thousand two hundred Facebook users who saw them use the combined method to beat purely social or purely network-based methods.. However, the CF-based technique fails to address cold start problems. Guo *et al.,*  [9] develop a lightweight referral scheme for privacy-conscious friends using a trust-based referral scheme for privacy-conscious friends. The model has demonstrated the practicability and privacy of the proposed scheme based on tracked preliminary results and security analysis.

Zheng, *et al.* [10] proposed a model to automatically infer users' modes of transportation, including driving, walking, bus riding, and cycling, from the GPS logs, using   supervised learning approaches. Hsu, *et al.,*  [11] propose a  friend recommendation system using collaborative and structural recommendation from friends with  weblog-based analysis of social networks [12]. Zhao,  *et al.,* [13] suggested a method for  finding social roles and status of people in online social networks by examining the  network structures. Chen *et al.,* [14] proposed a layered friendship model that talks about how the Mobile Social Networks (MSNs) are spreading in the face of the success of online social networks (OSNs) like Instagram. iPhones and Android phones. MSNs propagate existing OSNs by providing the ability to interact with new people who share his / her interests and by allowing a user to know when friends are nearby using the correlation between the friendship of the users with the possession of social charts, user profiles and mobility individualities. Zhou et al., [15] propose a  model for maintaining privacy in social network against attacks from the neighborhood". Even if the victim's self is preserved using conservative anonymization techniques.

## 3. PROBLEM DEFINITION AND SYSTEM ARCHITECTURE

The primary goal of any friend recommendation system in a social network is to offer with the most pertinent data to the user founded on their requirement or demand in order to advise friends. But these days in social networks there is an abundance of data leading to an overwhelming condition The main objective of this work is that since there are previously tons of morals and guidelines for friend recommendation in social network, we chose to continue with investigation of user personality/behavior and based on what the user want when we try to recommend friends for them. We have made use of 10 books that are distributed amongst the users in the dataset that are read by users and can help us recommend friends.

### 3.1. System Architecture

This has three vital modules: user modeling module, recommended object modeling module, and recommendation algorithm module. The recommendation system corresponds to the interest obligation essential in the user model along with that of the feature information in the recommendation object model, and while this is happening at the same time, the consistent recommendation algorithm is used to perform the computational model screening and the recommendation object that the user may have found and become attentive to found and then recommended to the user. If you then recommend, similar and closely related attitudes towards life will be sought. Here we have selected books to recommend to friends.

- **User Modeling**

It is characteristically recognized that information systems are becoming more involved than before and because of this, intellectual user interfaces are anticipated to recover user interaction with these numerous systems. Also, the exponential growth of the Internet makes it challenging for the users to cope with such huge amounts of accessible on-line data and can cause overload.

- **Recommended Object Modeling**

This component deals primarily with the feature information present. In machine learning and pattern recognition, a feature is defined as a distinct assessable property or distinguishing of an incidence that is being observed. Choosing explanatory, perceptive and self-governing features is a vital step for operative algorithms such as KNN Algorithm. Feature selection methods are mainly intended to lessen the number of contribution variables to those that are alleged to be most treasured to a model in order to predict the wanted target variable.

- **Recommendation Algorithm**

The recommendation algorithm is used to perform the calculation of the computational model and the recommendation object into which the user can be included is found and then recommended to the user.

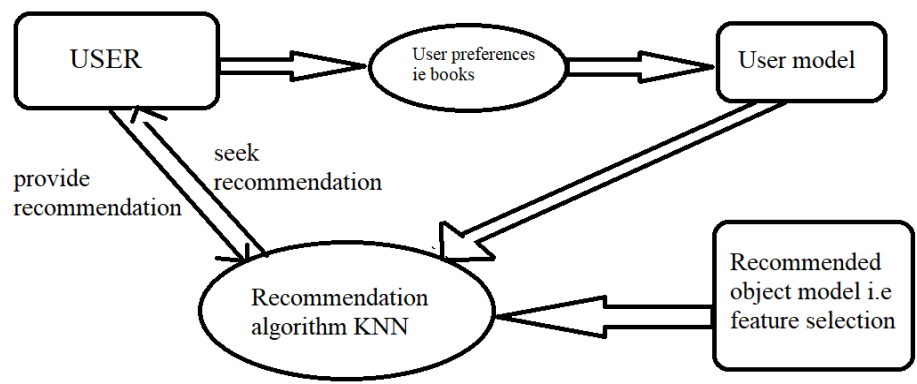

**Figure 1: System Architecture**

- **KNN Algorithm**

The KNN algorithm is known as the Supervised Machine Learning algorithm and it relies on labeled input data to learn a particular function that will produce appropriate output when the system is supplied with new unlabeled data. The KNN algorithm is carried out based on the $K$ neighboring values. $K$ is called a user-defined constant, and an unlabeled vector containing either a query or a test point is classified by assigning the label that is most repetitive or most persistent among the $K$ training patterns that are adjacent to that query point. The K-NN mode of operation can be explained using the algorithm:

- ➤ **Distance Metrics**

The distance metric is termed as an effective hyper-parameter through which we amount the distance between new test inputs and data feature values. We used the Euclidean distance to measure the distance between the new or unknown data point and the other neighbors. The nearest neighbor is then selected to classify the new data point. There are many benefits of which one benefit is that the distance between any two objects is not pretentious by the totaling of new objects to the analysis, which may be outliers. The Euclidian distance formula is given as follows where we find the distances between any given instances.

$$\text{Distance (d)} = \sqrt{(x-a)^2 + (y-b)^2} \qquad (1)$$

> **Choosing Optimal *K* value**

The *K* value determines or tells us the number of closest neighbors. First, we need to initialize a random *K*-value and start the calculation using Sqrt (*n*), where, *n* is the total number of data points. We then, derive a plot between of the error rate and *K* in a fixed range. Then we choose the *K* value having minimum error rate. Then, compute by choosing a small value of *K* and then find the error rates then chose the value of *K* which gives minimal error rate.

## 3.2. Data Preprocessing

- **Validation dataset**

A validation data set acts as a hybrid, meaning it is training data that is used for testing, but not as part of the low-level training, nor as part of the final test. Overfitting can be a big problem. In order to avoid overfitting, it is therefore necessary to have a validation data set in addition to the training and test data sets if organizational parameters need to be adjusted.

- **Splitting datasets**

The original idea is to split the data set T into two subsets - one subset that will be used for training, while the other subset will be left out and used to measure and evaluate the performance of the final model. The main purpose of cross-validation is to get a steady and safe approximation of model performance. We divide the data into training and test sets with a breakdown of 70 to 30, with 70 for training and 30 for testing.

- **Train Model**

Training a classification or regression model with a train model is a typical example of supervised machine learning. That means we need to provide a data set that contains historical data from which patterns can be learned.

## 3.3. Model Architecture

The data what the user has entered in the front end is stored in the file const.txt and this data is passed to the index.php file where this data is read and the required output is returned. The data inside const.txt file is sent to web_handler.py file and the consequence is received in the table procedure by the output variable. Since all the books are labelled as B0, B1 etc. we have to convert it into numbers first and then drop the user as it is a label. The dataset is the split into the ratio of 70:30 and the neighbors is set to 2 since it gives more correctness for training and testing the classifier. This value can also be changed to get dissimilar accuracy values.

- **Mapping**

Dictionaries are the mapping type built into Python. They are used to draw keys, which can be of any immutable type, into values, which can be of any type (heterogeneous), just like the elements of a list or a tuple. They are called associative collections because they associate a key with a value. Map () iterates through all the bases in the order given. During the iteration of the sequence, the specified callback () function is called for each element and then the returned value is made available in a new categorization. In the end, we return this new sequence of bad items. We then need to save the record values and the classifier in order to use run for the PHP in the front end.

# 4. EXPERIMENTAL RESULTS

## 4.1. Experimental Setup

The model was trained on a system having minimum 1 GHz but we recommend 2GHz or more GHz processor with minimum 32 GB, recommended 64 GB or more GB memory and GPU support of 8 GB or above. The model was developed with Python 3.6, a universal programming language at a high level. We can use Python for developing desktop GUI applications, web applications, and many websites. The Apache server is also used when the Apache HTTP server project is to be used to develop and maintain an open source HTTP server for modern functional schemes such as UNIX and Windows. The Apache server is set up to iterate through configuration records in which instructions are added to control its behavior. We have also used the Jupyter Notebook which is an open-source web application that permits us to make or create and share files that comprise quick code, comparisons, visualizations and narrative text. In total to running the code, it supplies code and output, together with notes that could prove useful in an editable document called a notebook. The database used is MYSQL.

There are mainly three classes of datasets used namely: User, Friend, Book. The dataset is the 'Facebook.csv' which we have extracted from Stanford snap Facebook Data which is referenced from Kaggle. This dataset contains a total of 4031 rows and 2 columns i.e. 4031R x 2C. We have added another column to this dataset 'book' which helps map the users to potential friends. The file is loaded, books is in the form of a list containing 10 books and the length of these books is stored in len_books variable. This file shows us the working for book dataset and gives the output where we see the number of books that have been read by a particular user

Here, we are using this function to randomly generate the books and also shuffling them. shuffle() is also used since it can randomize the items of a list. We have also made use of sets and lists. Set() is used to convert any iterable to sequence of iterable elements with discrete elements, Here a and b are taken as sets. We have then converted the sets into lists for easier calculations. Lists are said to be data structures in Python that can be mutable, or variable, methodical sequence of elements.

## 4.2. Pre-processing

This section gives the results we have obtained after preprocessing the dataset. It can be referenced by the Table 1 which shows output for our book dataset. It shows 3 columns which include necessary information for further classification. We implement this model by using KNN Classifier. It is hence imported and gives us the prediction output that is the accuracy of the model also for various values of $K$.

**Table 1: Result after preprocessing dataset**

| USER | FRIEND | BOOK |
|:---:|:---:|:---:|
| 1 | 0 | B7 |
| 0 | 1 | B7 |
| 1 | 0 | B9 |
| 0 | 1 | B9 |
| 1 | 0 | B4 |

| 0 | 1 | B4 |
|---|---|---|
| . | . | . |
| . | . | . |
| 7 | 0 | B9 |

- **Classification Rate/Accuracy**

We can calculate the accuracy scores for various values of k using the formula given. For the given dataset we also check the accuracy for various values to compare our accuracies.

$$Accuracy = \frac{Count*100}{(len\ (Y\_Test))}$$

- **Accuracy Comparison**

The Table 2 shows us how we obtain accuracy values for different values of $K$. The dataset is the split into the ratio of 70:30 and the neighbors is set to $K$ value to check for the accuracy after training and testing the classifier. We do the alpha and beta testing and after that we obtain the results as shown in the Table 2.

Table 2: Accuracy chart for different values of K

| K VALUE | ACCURACY OBTAINED |
|---|---|
| 2 | 75.125% |
| 3 | 83.342% |
| 4 | 87.876% |
| 5 | 90.534% |

# 5. CONCLUSIONs AND FUTURE SCOPE

Outlining a recommender system for a social network can prove hugely problematic as the things suggested here work less well. At this point, if a friend has been approved to a user and the user sends a friend request, the friend can definitely decline or delete the request if they don't care or want to keep a low profile. There are numerous social elements that play a role in creating an affiliation or bond between users. Therefore, we try to apply innumerable techniques in our recommendation system to improve the development of our recommendation system. We used the KNN algorithm for recommendation. The algorithm therefore does not take into account all neighboring points at the same time in order to bundle a single point, but iteratively only takes into account a closest point based on a selection method using a distance vector. Experimental results show that our model gives 90% accuracy.

Keeping our system up and running converts into a crucial task. So upkeep has to be shifted for the system to run in a smooth manner. Our recommendation accuracies can also go marginally mistaken in certain cases and we hope to develop a better accuracy score in future work. Bottlenecks arise while there is a search for neighbors, which are other users who have factually shown similar preferences to a given user, among huge user populations. We hope to correct these issues in further developments of improving our accuracies in future work related to this model.